\documentclass[a4paper,12pt]{amsart}
\usepackage{amssymb}
\usepackage{amscd}

\newcommand{\qft}{quantum Fourier transform\;}
\newcommand{\ct}{control/target inversion property\;}

\begin{document}
\title[inversion property]{Control/target inversion property on abelian groups}
\date{}
\author[M. Amini]{Massoud Amini}
\address{Department of Mathematics and Statistics\\University of Calgary\\ 2500 University Drive N.W., Calgary\\Alberta, Canada T2N 1N4\\
\linebreak
mamini@math.ucalgary.ca
\linebreak
Permanent address: Department of Mathematics\\Tarbiat Modarres University\\ B.O.Box 14115-175\\Tehran, Iran\\
amini@modares.ac.ir}
\thanks{This research was done while I was visiting University of Calgary, I would like to thank the Deaprtment of Mathematics and Statistics in U of C for their support. I am grateful to Professor Richard Cleve, for introducing me to this problem.}

\begin{abstract}
We show that the \qft on finite fields used to solve query problems is a special case of the usual \qft on finite abelian groups.
We show that the control/target inversion property holds in general. We apply this to get a sharp query complexity separation between
classical and quantum algorithms for a hidden homomorphism problem on finite abelian groups.
\end{abstract}

\maketitle

\section{Introduction}
One of the models which is used in checking the outperformance of quantum algorithms versus classical algorithms is the query model.
In this model, the input can only be accessed by means queries to a {\it black box}. Efficiency of computation then is measured by the
number of required queries. A famous example of query algorithm is Grover's algorithm [Gr] for searching a list of $n$ elements with $O(\sqrt{n})$ quantum queries.

In query complexity computation, one usually tries to find efficient quantum algorithms as well as lower bounds on the number of queries that any quantum or
classical algorithm needs. This lower bound or exact or bounded-error classical algorithms is used to check the outperformance of a given efficient quantum algorithm
over all possible classical counterparts. Probably the first instance of such an outperformance was demonstrated in the Deutsch algorithm [D], extended by Deutsch and Jozsa in [DJ].
The later solves an $(n+1)$-bit query problem using one query by a quantum algorithm with a lower bound of $\Omega (2^n)$ queries in exact classical solutions. Although it
turned out later that this problem could be solved using $O(1)$ queries with a bounded-error classical algorithm, The same query complexity separation has shown to exist between
quantum and bounded-error classical algorithms [BV]. This kind of separation has been pushed further in [BCW] in which a $2n$-query problem is presented that is solved by a quantum
algorithm using one query and has a lower bound of $\Omega(2^{\frac{n}{2}})$ in any bounded-error classical solution. The problem discussed in [BCW] is called the {\it hidden
linear structure} problem and is defined on a finite field $GF(2^n)$ (identified with $\{0,1\}^n$) as follows

\vspace{.3 cm}
{\bf Hidden Linear Structure Problem}. Let $\pi$ be a permutation on $GF(2^n)$ and $s\in GF(2^n)$. Define a black box $B_s$ on $GF(2^n)\times GF(2^n)$ by
$$B_s(x,y)=(x,\pi(y+sx))$$
Determine the value of $s$.

\vspace{.3 cm}
The quantum algorithm in [BCW] is based on a version of the \qft (QFT) on finite fields (a similar operation is used in [DH] to solve a shifted quadratic character problem).
The argument in [BCW] then proceeds using a \ct of the QFT. This is an intertwining property involving two linear operators defined by algebraic operations involving $s$.

In this paper we show that this is nothing but the usual \qft on the
abelian group $(GF(q),+)$ with respect to a special choice of the Fourier basis. Then we show that the \ct holds for a wide classes of
group homomorphisms on a general finite abelian groups. We use this to show that there is a sharp query complexity separation between
the bounded-error classical and exact quantum algorithms in solving a generalization of the linear structure problem in the context of abelian groups.
This problem could be called a hidden homomorphism problem and is stated as follows. Let $G$ be a finite (additive) abelian group and fix a Fourier
basis $\Lambda$ for the group algebra $\mathbb C\hat G$, let $Hom_\Lambda(G,G)$ be the set of all group homomorphsms on $G$ which are compatible with $\lambda$ (see section 2 for details),
then the problem is

\vspace{.3 cm}
{\bf Hidden Group Homomorphism Problem}. Let $\pi$ be a permutation on $G$, $a\in G$ and $\psi\in Hom_\Lambda (G,G)$. Define a black box $B_\psi$ on $G\times G$ by
$$B_\psi(x,y)=(x,\pi(y+\psi(x)))$$
Determine the value of $\psi(a)$.

\vspace{.3 cm}
When $G= (GF(q),+)$, $a=1$, and $\psi(x)=sx$ this problem reduces to the hidden structure problem.

In section $2$ we review the QFT on finite abelian groups. Our basic reference is [J]. In section $3$ we prove the \ct on groups.
Section $4$ is devoted to the quantum solution of the hidden group homomorphism problem and corresponding classical lower bounds.
In the section we present a variation of the \ct which leads to another generalization of the results of [BCW] to non commutative rings.

\section{The \qft on abelian groups}

Let $G$ be a finite abelian group. To emphesize that our group is abelian we use the addition as the group operation (this also helps to
avoid  any confusion when we later deal with the additive group of a finite field). Let $\mathcal H$ be a Hilbert space with the
orthonormal basis $\{|x\rangle: x\in G\}$, called the {\it standard basis} of $\mathcal H$. Indeed the group
algebra $\mathbb CG$ is a candidate for this Hilbert space. There is a natural action of $G$ on $\mathcal H$ by translation
$$x:|y\rangle\mapsto|x+y\rangle\quad (x,y\in G)$$
A {\it character} on $G$ is a nonzero group homomorphism
$\chi: G\to \mathbb T$, where $\mathbb T$ is the multiplicative group of the complex numbers of modulus 1. As each $x\in G$ has an
order dividing $n:=|G|$, the values $\chi(x)$ are $n$th roots of unity. The set $\hat G$ of all characters on $G$ is an abelian group
with respect to the pointwise multiplication and is called the {\it dual group} of $G$. It is well known that $|\hat G|=|G|=n$, and if
$\hat G=\{\chi_1,\dots,\chi_n\}$ then we have the Schur's orthogonality relations
$$\frac{1}{|G|}\sum_{x\in G} \chi_i(x)\overline{\chi_j(x)}=\delta_{ij},$$
for each $1\leq i,j\leq n$.

We prefer to index the elements of $\hat G$ by elements of $G$, so we write $\hat G=\{\chi_x: x\in G\}$. For each $x\in G$ cosider the
state
$$|\chi_x\rangle=\frac{1}{|G|}\sum_{y\in G} \overline{\chi_x(y)}|y\rangle,$$
then the above orthogonality relations imply that $\{|\chi_x\rangle: x\in G\}$ forms a orthonormal basis for $\mathcal H$, called the {\it Fourier
basis} of $\mathcal H$. This basis is translation invariant in the sense that
$$x|\chi_y\rangle=\chi_y(x)|\chi_y\rangle\quad (x,y\in G)$$
Also we may always assume that $\chi_x\chi_y=\chi_{x+y}$ and $\chi_0\cong 1$. Let $\psi:G\to G$ be a group homomorphism. We say that $\psi$ is compatible
with the Fourier basis of $G$ if
$$\chi_y(\psi(z))=\chi_{\psi(y)}(z)\quad (y,z\in G)$$
Given a Fourier basis $\Lambda$ (that is a given choice of the indexing $\hat G$ with $G$) we denote the set of all homomorphisms $\psi$ of $G$ compatible
with $\Lambda$ by $Hom_\Lambda (G,G)$. On any finite abelian group $G$ we have a family of such homomorphisms constructed using the structure theorem for $G$.
Every finite abelian group $G$ is isomorphic to the Cartesian product of cyclic groups, say $G=\prod_{1\leq j\leq k} \mathbb Z_{m_j}$. For each $x=(x_1, \dots, x_k)\in G$
with $x_j\in \mathbb Z_{m_j}$, we have the character
$$\chi_x(y)=\prod_{1\leq j\leq k} \omega_j^{x_jy_j}\quad (y=(y_1,\dots,y_k)\in G)$$
where $\omega_j=e^{\frac{2\pi i}{m_j}}$ and the product $x_jy_j$ is calculated $(mod \, m_j)$. Then $\Lambda=\{\chi_x: x\in G\}$ is a Fourier basis and for each
$s=(s_1,\dots,s_k)\in G$ defines a homomorphism $\psi_s$ by
$$\psi_s(y)=(s_1y_1,\dots, s_ky_k) \quad (y=(y_1,\dots,y_k)\in G)$$
which is clearly compatible with $\Lambda$. Here again the products $s_jy_j$ are defined $(mod \, m_j)$.

The {\it \qft} on $G$ is the unitary operator
$F_G:\mathcal H\to\mathcal H$ defined by
$$|x\rangle\mapsto \frac{1}{\sqrt{|G|}}\sum_{y\in G} \chi_x(y)|y\rangle\quad(x,y\in G)$$
Note that one can extend this map by linearity on $\mathcal H$ and the fact that it is unitary follows from Pontryagin duality for abeliab
groups [J]. Two classical examples are $G=\mathbb Z_m$ where
$$\chi_k(\ell)=e^{2\pi ik/m}\quad k,\ell=0,\dots,m-1$$
and $G=\{0,1\}^n$ where
$$\chi_x(y)=(-1)^{x.y}\quad (x,y\in\{0,1\}^n)$$
in which $F_G$ is the usual discrete Fourier transform $DFT_m$ on $\mathbb Z_m$ and the Hadamard transform $H_n$, respectively. Another example
would be the additive group of any finite field $GF(q)$, which is discussed in details in the next section.

\section{main result}
Let $G$ be an (additive) abelian group and $\mathcal H=\mathbb CG$. Let $\Lambda$ be a Fourier basis for $\mathcal H$.
To each homomorphism $\psi\in Hom_\Lambda(G,G)$, there corresponds two operators on $\mathcal H\otimes\mathcal H$ defined by
\begin{align*} &A_\psi:|x\rangle|y\rangle\mapsto |x\rangle|y+\psi(x)\rangle\\
&B_\psi:|x\rangle|y\rangle\mapsto |x+\psi(y)\rangle|y\rangle
\end{align*}
We say that a unitary operator $U$ on $\mathcal H$ satisfies the \ct at $\psi$ if
$$(U^{\dagger}\otimes U)A_\psi(U\otimes U^{\dagger})=B_\psi$$
\vspace{.3 cm}
{\bf Theorem 1 (Main Result).} Let $G$ be a finite abelian group and $\psi$ be a group homomorphism on $G$. Choose a Fourier basis
of $\mathcal H=\mathbb CG$, then for each $\psi\in Hom_\Lambda(G,G)$,  the \qft $F_G$ satisfies the \ct at $\psi$.

{\bf Proof.} Let $n=|G|$, for $x\in G$ put
$$|F_x\rangle=F_G|x\rangle= \frac{1}{\sqrt{n}}\sum_{y\in G} \chi_x(y)|y\rangle$$
and $P_x|y\rangle=|x+y\rangle$. Then
\begin{align*}
P_y|F_{-x}\rangle&=\frac{1}{\sqrt{n}}\sum_{z\in G} \chi_{-x}(z)|y+z\rangle\\
&=\frac{1}{\sqrt{n}}\sum_{z\in G} \chi_{-x}(z-y)|z\rangle\\
&=\frac{1}{\sqrt{n}}\sum_{z\in G} \chi_{-x}(-y)\chi_{-x}(z)|z\rangle\\
&=\chi_{-x}(-y)|F_{-x}\rangle=\chi_{x}(y)|F_{-x}\rangle
\end{align*}
Therefore
\begin{align*}
(F_G^{\dagger}\otimes F_G)A_\psi(F_G\otimes F_G^{\dagger})|x\rangle|y\rangle\\
&=(F_G^{\dagger}\otimes F_G)A_\psi\big(\frac{1}{\sqrt{n}}\sum_{z\in G} \chi_{x}(z)|z\rangle|F_{-y}\rangle\big)\\
&=(F_G^{\dagger}\otimes F_G)\big(\frac{1}{\sqrt{n}}\sum_{z\in G} \chi_{x}(z)|z\rangle P_{\psi(z)}|F_{-y}\rangle\big)\\
&=(F_G^{\dagger}\otimes F_G)\big(\frac{1}{\sqrt{n}}\sum_{z\in G} \chi_{x}(z)|z\rangle\chi_{y}(\psi(z))|F_{-y}\rangle\big)\\
&=(F_G^{\dagger}\otimes F_G)\big(\frac{1}{\sqrt{n}}\sum_{z\in G} \chi_{x}(z)\chi_{\psi(y)}(z)|z\rangle|F_{-y}\rangle\big)\\
&=(F_G^{\dagger}\otimes F_G)\big(\frac{1}{\sqrt{n}}\sum_{z\in G} \chi_{x+\psi(y)}(z)|z\rangle|F_{-y}\rangle\big)\\
&=(F_G^{\dagger}\otimes F_G)|F_{x+\psi(y)}(z).|F_{-y}\rangle\\
&=|x+\psi(y)\rangle|y\rangle=B_\psi|x\rangle|y\rangle.QED
\end{align*}
\vspace{.3 cm}

As a basic example let us consider the main example of [BCW]. Let $GF(q)$ be the finite field with $q=p^m$ elements and fix an irreducible
polynomial $f(Z)=Z^m-\sum_{i=0}^{m-1} a_iZ^i$ over $GF(p)$, and let $<f>$ be the ideal generated by $f$, then
$$GF(q)\simeq\frac{GF(p)[Z]}{<f>}$$
Fix a nonzero linear map $\varphi:GF(q)\to GF(p)$ and define the \qft $F_{q,\varphi}:\mathbb CGF(q)\to\mathbb CGF(q)$ by
$$F_{q,\varphi}:|x\rangle\mapsto \frac{1}{\sqrt{q}}\sum_{y\in GF(q)}e^{2\pi i\varphi(xy)/p}|y\rangle,$$
extended by linearity. Then the additive group $G:=(GF(q),+)$ is an abelian group and for each $x\in G$
$$\chi_x(y)=e^{2\pi i\varphi(xy)/p}\quad(y\in G)$$
defines a character on $G$. Also we have the orthogonality relations
$$\frac{1}{q}\sum_{x\in G} \chi_y(x)\overline{\chi_z(x)}=\frac{1}{q}\sum_{x\in G} e^{2\pi i\varphi(x(y-z))/p}=\delta_{yz},$$
(see the proof of [BCW, Theorem 1]). Also if $\chi_x=\chi_y$, then $e^{2\pi i\varphi(xz)/p}=e^{2\pi i\varphi(yz)/p}$, for each $z\in G$.
Since the range of $\varphi$ is in $GF(p)=\mathbb Z_p$ and the analytic map $\omega\mapsto exp(\omega)$ is one-to-one in the strip
$\{\omega\in \mathbb C: 0\leq Im(\omega)<2\pi\}$, we get $\varphi(xz)=\varphi(yz)$, for each $z\in G$. If $x\neq y$, then we have $q$
distinct elements $z(x-y)$ in $ker(\varphi)$, which means that $ker(\varphi)=G$, i.e. $\varphi=0$, which is a contradiction. Hence $x=y$,
that is $\{\chi_x: x\in G\}$ is a complete set of Fourier basis elements for $G$. Now it is clear that, with respect to this basis,
$F_G=F_{q,\varphi}$. Next let $s\in G$ be any nonzero element and define $\psi_s(x)=sx\quad (x\in G)$. This is clearly a group homomorphism
of $G$ which is compatible with the above Fourier basis, namely

\begin{align*}
\chi_{\psi_s(y)}(z)&=exp\big(2\pi i \varphi((sy)z)/p\big)=exp\big(2\pi i \varphi((sz)y)/p\big)\\
&=exp\big(2\pi i \varphi(\psi_s(z)y)/p\big)=exp\big(2\pi i \varphi(y\psi_s(z))/p\big)=\chi_y(\psi_s(z))
\end{align*}

In particular Theorem 1 of [BCW] is an special case of our main theorem. Also note that for the additive group $G$ of a commutative ring all
the above observations are valid except that $\{\chi_x: x\in G\}$ is not necessarily a complete set of Fourier basis elements for $G$
(we need commutativity of the ring in the second equality of the second line of the above calculation to show
that $\psi_s$ is compatible with the Fourier basis). In the last section of [BCW] there is a version of the \ct for the ring
of $m\times m$ matrices over a commutative ring $R$. This is again a special case of a minor modification of the above theorem.
Consider a pair $(\psi,\varphi)$ of homomorphisms of $G$ such that $\psi\circ\varphi =\varphi\circ\psi$. We say that $(\psi,\varphi)$ is
{\it compatible} with a given Fourier basis $\Lambda$ of $G$ if
$$\chi_y(\psi(z))=\chi_{\varphi(y)}(z)\quad (y,z\in G)$$
We denote the set of all such pairs by $Hom_{\Lambda,\Lambda}(G,G)$. We say that a unitary operator $U$ on
$\mathcal H$ satisfies the \ct at $(\psi,\varphi)$ if
$$(U^{\dagger}\otimes U)A_\psi(U\otimes U^{\dagger})=B_\varphi$$
Then a slight modification of the proof of Theorem 1 shows that

\vspace{.3 cm}
{\bf Theorem 2.} Let $G$ be a finite abelian group and choose a Fourier
basis of $\Lambda$ of $\mathcal H=\mathbb CG$, then for each $(\psi,\varphi)\in Hom_{\Lambda,\Lambda}(G,G)$, the \qft $F_G$ satisfies the \ct at $(\psi,\varphi)$.

\vspace{.3 cm}
Now in section $4$ of [BCW], we are dealing with a ring $R$ with QFT $F_R$ for which a QFT $F_{R,m}$ is defined on the ring $R^{m\times m}$
of $m\times m$ matrices over $R$ via tensor product. It is clear that if $F_R$ is the QFT on the additive group $G=(R,+)$, then
$F_{R,m}$ is the QFT on the product group $G^{m^2}$ (which is the additive group of the ring $R^{m\times m}$). The two group homomorphisms of
$G^{m^2}$ are then $\psi(X)=SX$ and $\varphi(X)=XS$\quad $(X\in R^{m\times m})$, where $S$ is an element of $R^{m\times m}$. Now
with the natural choice of the Fourier basis for $G=(R,+)$ we would have
$$\chi_{y}(sz)=\chi_{ys}(z)\quad (s,y,z\in R)$$
Define the Fourier basis of $G^{m^2}$ by
$$\chi_Y(Z)=\prod_{i,j=1}^{m} \chi_{y_{ij}}(z_{ji})\quad (Y=[y_{ij}], Z=[z_{ij}]\in R^{m\times m})$$
Then for each $S,Y,Z\in R^{m\times m}$, we have
\begin{align*}
\chi_Y(SZ)& =\prod_{i,j=1}^{m} \chi_{y_{ij}}((SZ)_{ji})=\prod_{i,j=1}^{m} \chi_{y_{ij}}(\sum_{k=1}^{n}s_{jk}z_{ki})\\
&=\prod_{i,j,k=1}^{m} \chi_{y_{ij}}( s_{jk}z_{ki})=\prod_{i,j,k=1}^{m} \chi_{y_{ij}s_{jk}}(z_{ki})\\
&=\prod_{i,k=1}^{m} \chi_{\sum_{j=1}^{m} y_{ij}s_{jk}}(z_{ki})=\prod_{i,k=1}^{m} \chi_{(YS)_{ik}}((Z)_{ki})=\chi_{YS}(Z)
\end{align*}
Therefore the \ct presented in section $4$ of [BCW] follows from Theorem $2$ above.

\section{the hidden homomorphism problem}

For a finite (additive) abelian group $G$ let $a\in G$ be a fixed element (usually the generator of $G$, when $G$ is cyclic),
$\pi$ be an arbitrary permutation of elements of $G$, and for a fixed Fourier basis
$\Lambda:=\{\chi_x: x\in G\}$ of $\mathcal H=\mathbb CG$, let $\psi\in Hom(G,G)$ be a homomorphism of $G$ compatible with $\Lambda$, then
the {\it hidden homomorphism problem} on $G$ is as follows: Given a black-box performing the unitary transformation that maps $|x\rangle|y\rangle$ to
$|x\rangle|\pi(y+\psi(x)\rangle$, find $\psi(a)$. In this section we show that, using the QFT, a single query is sufficient to solve the problem exactly,
where as in the classical case, even for cyclic groups, $\Omega(|G|^{\frac{1}{2}})$ queries are needed to solve the problem with bounded error.

\vspace{.3 cm}
{\bf Theorem 3.} On any finite abelian group $G$, performing $F_G$ and $F_G^{\dagger}$, a single query is sufficient to solve the hidden
homomorphism problem exactly.

{\bf Proof.} Consider the unitary transformation
$$U_\pi: |y\rangle\mapsto |\pi(y)\rangle$$
implementing $\pi$ and recal that
$$A_\psi: |x\rangle|y\rangle\mapsto |x\rangle|y+\psi(x)\rangle,$$
then the black-box is implemented by
$$U_{\pi,\psi}:=(I\otimes U_\pi)A_\psi : |x\rangle |y\rangle \mapsto |x\rangle|\pi(y+\psi(x)\rangle.$$
To perform the quantum procedure, first initialize the state of two $G$-valued registers to $|0\rangle|a\rangle$, where $0$ is the identity of $G$.
Then perform the following consecutive operations: apply $F_G\otimes F_G^{\dagger}$, then query the black-box and apply
$F_G^{\dagger}\otimes F_G$. Finally measure the first register. The states of the two registers during the exacution of this algorithm
is as follows:

\begin{align*}
&|0\rangle|a\rangle \stackrel{F_G\otimes F_G^{\dagger}} {\longrightarrow} F_G|0\rangle F_G^{\dagger}|a\rangle
\stackrel{U_{\pi,\psi}}{\longrightarrow} (I\otimes U_\pi)A_\psi (F_G\otimes F_G^{\dagger})|0\rangle|a\rangle\\
&=(I\otimes U_\pi)(F_G\otimes F_G^{\dagger})B_\psi|0\rangle|a\rangle=(I\otimes U_\pi)(F_G\otimes F_G^{\dagger})|\psi(a)\rangle|a\rangle\\
&=F_G|\psi(a) \rangle U_\pi F_G^{\dagger}|a\rangle \stackrel{F_G^{\dagger}\otimes F_G} {\longrightarrow} |\psi(a) \rangle (F_G U_\pi F_G^{\dagger})|a\rangle
\end{align*}

Now measuring the first register gives $|\psi(a)\rangle$. QED

\vspace{.3 cm}
{\bf Theorem 4.} On any finite cyclic group $G$ of prime order,\,$\Omega(|G|^{\frac{1}{2}})$  queries are necessary to solve the hidden
homomorphism problem within probability error $\frac{1}{2}$.

{\bf Proof.} By an argument similar to [BCW, Theorem 3] we may deterministic algorithms with probabilistic input data (here we put
$a=1$, the generator of $G$). Set $\psi\in Hom(G,G)$ and $\pi$ randomely with uniform distribution. After $k$ (distinct) queries
$(x_1,y_1),\dots,(x_k,y_k)$, if there are two indecies $i\neq j$ such that $\pi(y_i+\psi(x_i))=\pi(y_j+\psi(x_j))$, then, as $\pi$ is
one-to-one, $y_i-y_j=\psi(x_j-x_i)=(x_j-x_i)\psi(1)$, and $\psi(1)$ is uniquely determined, otherwise we have $|G|-k(k-1)/2$ possibilities
for $\psi(1)$ which are equally likely. A simple argument shows that the probability of a collision occurring at the $k$th query is at most
$\frac{k-1}{|G|-(k-1)(k-2)/2}$. Therefore the probability of a collision occcurring in the first $m$ queries is bounded above by
$$\sum_{k=1}^{m} \frac{k-1}{|G|-(k-1)(k-2)/2} \leq \sum_{k=1}^{m} \frac{2k}{2|G|-k^2}\leq \frac{m^2}{2|G|-m^2},$$
this being at least $\frac{1}{2}$, implies that $m\geq (\frac{2}{3}|G|)^{\frac{1}{2}}$. QED

\end{document}